\documentclass[aps,pra,english,nofootinbib,
floatfix,twocolumn]{revtex4-2}

\usepackage{amsfonts,amsmath,amssymb,mathrsfs,mathtools,bm}
\usepackage{graphicx}
\graphicspath{{figures/}}
\usepackage[utf8]{inputenc}
\usepackage[colorlinks=true,citecolor=blue,linkcolor=blue,urlcolor=blue]{hyperref}
\usepackage{babel}
\usepackage{color}

\newcommand\be{\begin{equation}}
\newcommand\ee{\end{equation}}
\newcommand\bea{\begin{eqnarray}}
\newcommand\eea{\end{eqnarray}}

\def\Tr{\operatorname{Tr}}

\newcommand{\Kry}{\mathscr K}

\begin{document}

\title{Krylov Distribution and Spectral 
Convergence of Quantum Fisher Information}
 
\author{M.~Alishahiha}
\affiliation{School of Quantum Physics and Matter\\ Institute for Research in Fundamental Sciences (IPM),\\ P.O. Box 19395-5531, Tehran, Iran}

\author{F.~T.~Tabesh}
\affiliation{School of Quantum Physics and Matter\\ Institute for Research in Fundamental Sciences (IPM),\\ P.O. Box 19395-5531, Tehran, Iran}
     
\author{M.~J.~Vasli}
\affiliation{School of Quantum Physics and Matter\\ Institute for Research in Fundamental Sciences (IPM),\\ P.O. Box 19395-5531, Tehran, Iran}
\begin{abstract}
We establish a direct correspondence between the Krylov hierarchy of quantum Fisher information (QFI) and a Krylov-resolved decomposition of the symmetric logarithmic derivative (SLD). Building on the existing Krylov formulation of QFI, we show that the metrological information resolved at each Krylov level is exactly encoded in the incremental gain of the Lanczos hierarchy, giving the Krylov distribution a direct information-theoretic interpretation. A dual spectral formulation distinguishes the Hilbert--Schmidt geometry used to construct the computational Lanczos hierarchy from the state-dependent geometry in which the SLD and QFI are naturally defined, and relates their associated Liouville measures. This distinction clarifies the origin of different convergence regimes: we recover exponential convergence for gapped spectra and derive an exact 
algebraic hard-edge convergence
law for the Jacobi class, including the 
exact finite-order error and the marginal 
case separating finite from divergent QFI.
\end{abstract}

\maketitle

\section{Introduction}

Quantum Fisher information (QFI) quantifies the sensitivity of a quantum state to infinitesimal unitary transformations, capturing its distinguishability under small parameter variations. Originally central to quantum metrology, where it sets the ultimate precision limit via the quantum Cram\'er--Rao bound~\cite{Helstrom1969,Braunstein1994,Paris2009}, QFI has since emerged as a fundamental probe across many-body physics. It witnesses 
multipartite entanglement~\cite{Toth2014,Hyllus2012,Pezz2009}, signals quantum criticality~\cite{Hauke2016,Campos,Zanardi}, characterizes nonlocality~\cite{Niezgoda}, 
and probes quantum 
geometry~\cite{Lambert}. In parallel, Krylov-space methods have become a standard language for operator-growth dynamics~\cite{Parker2019,Rabinovici2022}. Further QFI applications include quantum thermodynamics, quantum speed limits~\cite{Liu2019}, and non-Markovian dynamics~\cite{Scandi}.

This broad relevance creates a fundamental computational challenge: evaluating QFI requires resolving a state-dependent response problem in an operator space whose dimension grows quadratically with the Hilbert-space dimension. For a differentiable family of states $\rho_\theta$, the symmetric logarithmic derivative (SLD) is defined implicitly by
\begin{equation}
B\equiv\partial_\theta\rho_\theta
=
\mathcal K_\rho(L),
\qquad
\mathcal K_\rho(X)=\frac12\{\rho,X\},
\label{eq:intro-AB}
\end{equation}
where $B$ is the tangent operator. For unitary parameter encoding, $B=i[\rho,H]$, and the QFI is the state-dependent norm of the SLD,
\begin{equation}
\mathcal F=\Tr(\rho L^2).
\end{equation}
Evaluating QFI is therefore equivalent to solving the linear-response problem
\begin{equation}
A x=b,
\label{eq:linear-system}
\end{equation}
with identifications
\begin{equation}
A=\mathcal K_\rho,\qquad x=L,\qquad b=B .
\end{equation}
The difficulty lies in the fact that $A$ acts in Liouville space, whose dimension is the square of the Hilbert-space dimension. Direct construction or inversion of this operator rapidly becomes prohibitive in interacting many-body systems.

Krylov-subspace methods provide a natural alternative by avoiding explicit resolution of the full operator space. Instead of solving Eq.~\eqref{eq:linear-system} globally, the solution is approximated in the nested sequence
\begin{equation}
\mathcal K_n(A,B)={\operatorname{span}}\{B,AB,A^2B,\dots,A^{n-1}B\},
\end{equation}
for $n$ much smaller than the Liouville dimension. The Lanczos algorithm constructs an orthonormal basis of this subspace and generates a tridiagonal representation of $A$, reducing the original problem to a sequence of small projected problems~\cite{lanczos1950,park1986,haydock1980,saad2003iterative,viswanath1994,Druskin1989,dkz09}. Krylov shadow tomography established the monotone hierarchy of nonpolynomial QFI lower bounds, its variational interpretation, exact recovery at Krylov saturation, and a shifted-Hankel moment representation~\cite{Zhang:2025wqj}; Refs.~\cite{Nandy:2024evd,Rabinovici:2025otw} review the broader setting of Krylov-space dynamics.

However, an important structural question remains: what is the physical meaning of the Krylov hierarchy itself? The Krylov construction involves two distinct geometries. The algebraic Krylov space is determined only by $\mathcal K_\rho$ and $B$, but its orthogonal decomposition depends on the choice of inner product. The standard implementation considered here uses Hilbert--Schmidt Lanczos geometry, while the QFI is naturally measured in the state-dependent energy geometry, which also governs the decomposition of the SLD into information-carrying components.

Building on this established hierarchy, we make its level-by-level information content explicit by using the two geometries as complementary representations of the same finite-order response. Specifically, if $J_n$ denotes the Hilbert--Schmidt Lanczos Jacobi matrix and $\ell_k$ are the coefficients of the exact SLD in the filtration-adapted state-orthonormal Krylov basis, then
\begin{equation}
\mathcal F_n
=
\|B\|_{\mathrm{HS}}^2
e_0^\dagger J_n^{-1}e_0
=
\sum_{k=0}^{n-1}|\ell_k|^2,
\label{eq:intro-central}
\end{equation}
where $e_0=(1,0,\ldots,0)^{\mathsf T}$ and $\|B\|_{\mathrm{HS}}$ is the Hilbert--Schmidt norm of the seed operator $B$.

Equation~\eqref{eq:intro-central} is the main level-resolved identity used in this work. Its role is to turn the established computational QFI hierarchy into the cumulative distribution function of a Krylov information distribution,
\begin{equation}
p_k
=\frac{\mathcal F_{k+1}-\mathcal F_k}{\mathcal F}.
\end{equation}
Consequently, the convergence of QFI approximations acquires a direct physical interpretation: the truncation error is not merely a numerical residual but the unresolved probability weight residing in higher Krylov directions. The first moment of this distribution defines an intrinsic information depth that measures how many Krylov directions are required to resolve the metrological response.

This dual structure also has a spectral interpretation. The Hilbert--Schmidt and state-dependent formulations generate two related spectral measures on the active spectrum of $\mathcal K_\rho$. The former represents the QFI as a first inverse spectral moment and governs the Lanczos hierarchy, while the latter represents the same quantity through the state-metric SLD norm. These measures are connected by a Christoffel transformation: they differ only in spectral weighting while assigning the same physical QFI contribution to each active mode.

The dual spectral formulation enables a unified description of Krylov convergence in both regular and singular regimes. In the presence of a spectral gap, the standard conjugate-gradient interpretation is recovered. More importantly, the framework also captures hard-edge spectra, where the support of the kernel measure extends to zero. For the exactly solvable Jacobi class
\begin{equation}
d\mu_\alpha=(\alpha+1)\lambda^\alpha d\lambda,
\qquad
\lambda\in[0,1],
\end{equation}
with $\alpha>1$, the finite-order error is
\begin{equation}
1-\frac{\mathcal F_n}{\mathcal F}
=
\left[
\frac{\Gamma(\alpha)\Gamma(n+1)}
{\Gamma(n+\alpha)}
\right]^2
\sim
\Gamma(\alpha)^2 n^{-2(\alpha-1)} .
\label{eq:intro-hard-edge}
\end{equation}
This provides an analytically controlled benchmark for the singular regime, while more general hard-edge distributions require additional regularity conditions to determine the asymptotic exponent.

We further demonstrate the physical content of the dual Krylov framework through a matrix-free numerical implementation for a mixed-field Ising system. Working in the eigenbasis of the density matrix, the QFI kernel becomes a diagonal spectral problem, allowing the Lanczos recursion to be performed without constructing the Liouville-space superoperator. The numerical results show that the Krylov information distribution is robust across an ensemble of random full-rank states and that the QFI can be resolved to high precision in a small effective Krylov sector compared with the exponentially larger compressed spectral space. For example, in a representative $N=12$ system with compressed spectral dimension $8\,386\,560$, the QFI reaches a relative resolution of $10^{-15}$ after only
approximately $4500$ Krylov directions, demonstrating the remarkable compressibility of the metrological response in Krylov space.

The paper is organized as follows. Section~\ref{sec:QFI} introduces the QFI Krylov distribution and derives its cumulative, incremental, and tail relations. Section~\ref{sec:spectral} develops the dual spectral formulation, moment representations, and convergence theory in gapped and hard-edge regimes. Section~\ref{sec:applications} presents the matrix-free numerical validation and ensemble-averaged Krylov analysis. Section~\ref{sec:conclusions} summarizes the main results. Technical derivations and details of the two Krylov geometries are provided in the appendices.

\section{Quantum Fisher Information and Krylov Distribution}
\label{sec:QFI}

The symmetric logarithmic derivative is not only the solution of an operator-space linear equation; it is also a response operator whose state-dependent norm gives the quantum Fisher information. Resolving this norm within the cyclic space generated by the tangent operator therefore provides a natural way to identify how different Krylov directions contribute to the total metrological sensitivity.

The established hierarchy of Krylov QFI approximations can be interpreted as successive state-metric projections of the exact SLD~\cite{Zhang:2025wqj}. Here we give that hierarchy a filtration-resolved information interpretation. The construction is the operator-space counterpart of the resolvent Krylov distribution introduced in Ref.~\cite{Alishahiha:2026fnu}\footnote{See also~\cite{Lunt:2025dcc} 
for related ideas on operator spreading.}
: instead of resolving a dressed state in Hilbert space, we resolve the SLD, which is the inverse response of the state to the tangent $B=\partial_\theta\rho_\theta$.

Let $d_0$ denote the saturation dimension of the cyclic space generated by $B$. We use a \emph{filtration-adapted} state-orthonormal sequence $\{v_0,\ldots,v_{d_0-1}\}$, defined by
\begin{equation}
 \operatorname{span}\{v_0,\ldots,v_k\}=\Kry_{k+1},
 \qquad
 v_k\in\Kry_{k+1}\cap\Kry_k^{\perp_\rho}.
 \label{eq:adapted-state-basis}
\end{equation}
Here $\Kry_0=\{0\}$ and orthogonality 
is in the state metric (see Appendix~\ref{app:Krylov-review}). This nested condition fixes each level up to a phase (before saturation) and is essential: an arbitrary state-orthonormal rotation of the saturated cyclic space would change the individual level weights. The exact SLD can then be expanded as
\begin{equation}
 L=\sum_{k=0}^{d_0-1}\ell_k v_k,
 \qquad
 \ell_k=\langle v_k,L\rangle_\rho=(v_k,B)_0 .
 \label{eq:qfi-L-expansion}
\end{equation}
The second equality follows from the SLD equation $\mathcal K_\rho L=B$ and expresses each coefficient directly as an overlap with the physical tangent operator. It is therefore not necessary to apply $\mathcal K_\rho^{-1}$ in order to assign a Krylov-resolved weight to the response.

Since the saturated Krylov space contains the SLD on the active operator subspace, Parseval's identity gives
\begin{equation}
 \mathcal F
 =\langle L,L\rangle_\rho
 =\sum_{k=0}^{d_0-1}|\ell_k|^2 .
 \label{eq:qfi-parseval}
\end{equation}
This decomposition motivates the normalized weights
\begin{equation}
 p_k=\frac{|\ell_k|^2}{\mathcal F},
 \qquad
 \sum_{k=0}^{d_0-1}p_k=1,
 \qquad
 \mathcal D=\sum_{k=0}^{d_0-1}k\,p_k .
 \label{eq:qfi-pk}
\end{equation}
The probability $p_k$ is the fraction of the total QFI carried by the $k$-th metric-orthogonal Krylov direction. Accordingly, $\mathcal D$  measures the mean depth over which the metrological response is resolved.

This interpretation differs from the usual dynamical Krylov complexity. Here the Krylov sequence is generated by the positive anticommutator superoperator $\mathcal K_\rho$, and the SLD is a static inverse-response operator. Thus $\mathcal D$ should be understood as an information-resolution or resolvent depth. A small value indicates that the QFI is concentrated in response directions close to the tangent operator $B$, whereas a large value signals that a substantial fraction of the sensitivity is distributed over more deeply generated directions.

The $n$-dimensional approximation is the state-metric projection
\begin{equation}
 L_n=\sum_{k=0}^{n-1}\ell_k v_k .
 \label{eq:qfi-Ln-projection}
\end{equation}
Its QFI content is
\begin{equation}
 \mathcal F_n
 =\langle L_n,L_n\rangle_\rho
 =\sum_{k=0}^{n-1}|\ell_k|^2
 =\mathcal F\sum_{k=0}^{n-1}p_k .
 \label{eq:qfi-Fn-cumulative}
\end{equation}
The Krylov hierarchy is therefore the cumulative distribution function of the QFI weights~\cite{Zhang:2025wqj}:
\begin{equation}
 0=\mathcal F_0
 \leq\mathcal F_1
 \leq\cdots
 \leq\mathcal F_{d_0}
 =\mathcal F .
 \label{eq:qfi-monotone}
\end{equation}
Each Krylov enlargement resolves an additional nonnegative contribution to the metrological response.

Projection orthogonality also gives the exact Pythagorean relation
\begin{equation}
 \mathcal F-\mathcal F_n
 =\|L-L_n\|_\rho^2
 =\sum_{k=n}^{d_0-1}|\ell_k|^2 .
 \label{eq:exact-error}
\end{equation}
Consequently, the relative error is precisely the unresolved tail of the distribution,
\begin{equation}
 1-\frac{\mathcal F_n}{\mathcal F}
 =\sum_{k=n}^{d_0-1}p_k .
 \label{eq:qfi-tail}
\end{equation}
This identity is exact at every finite Krylov order. It is stronger than an asymptotic convergence estimate because it identifies the error with a specific portion of the SLD norm.

A general spectrum-independent bound follows immediately from the first moment of the distribution. Since $k\geq n$ throughout the unresolved tail,
\begin{equation}
 0\leq
 1-\frac{\mathcal F_n}{\mathcal F}
 \leq\frac{\mathcal D}{n}.
 \label{eq:D-tail-bound}
\end{equation}
This bound does not determine the asymptotic convergence law, which depends on the spectral measure studied in Sec.~\ref{sec:spectral}. It instead provides a nonasymptotic structural guarantee: for $n\geq1$, once the Krylov dimension is large compared with the information depth $\mathcal D$, only a small fraction of the QFI can remain unresolved. Because the exact $\mathcal D$ contains the full tail, it is not by itself an a priori stopping criterion in a truncated computation.

The Krylov distribution itself has an equivalent cumulative interpretation. Using Eq.~\eqref{eq:qfi-tail} and interchanging the finite sums gives
\begin{equation}
 \mathcal D
 =\sum_{n=1}^{d_0-1}
 \left(
 1-\frac{\mathcal F_n}{\mathcal F}
 \right).
 \label{eq:D-area}
\end{equation}
Thus $\mathcal D$ is the discrete area under the relative-error curve. It measures not only the location of the QFI distribution but also the aggregate number of Krylov steps over which metrological information remains unresolved. Consequently, $\mathcal D$ provides a global measure of the computational effort required to resolve the metrological response.

Although the coefficients $\ell_k$ are defined through the state-orthonormal representation, neither the exact SLD nor a second state-metric Lanczos recursion is required to obtain them. Let $J_n$ denote the tridiagonal matrix generated by the standard Hilbert--Schmidt Lanczos procedure applied to $\mathcal K_\rho$ with seed $B$, and let $\beta_0=\|B\|_{\mathrm{HS}}$. The central finite-order identity is
\begin{equation}
 \mathcal F_n
 =\beta_0^2 e_0^\dagger J_n^{-1}e_0
 =\sum_{k=0}^{n-1}|\ell_k|^2 .
 \label{eq:qfi-central-main}
\end{equation}
The first expression is the computational QFI hierarchy, while the second is the cumulative state-metric weight of the exact SLD. Their equality shows that the Hilbert--Schmidt Galerkin calculation and the information-geometric projection are two representations of the same finite-order response.

It follows in particular that the contribution of a single Krylov level is the gain obtained by enlarging the computational space by one dimension:
\begin{equation}
 |\ell_k|^2
 =\mathcal F_{k+1}-\mathcal F_k,
 \qquad
 p_k
 =\frac{\mathcal F_{k+1}-\mathcal F_k}{\mathcal F}.
 \label{eq:qfi-increments}
\end{equation}
A Hilbert--Schmidt Lanczos run therefore yields the monotone hierarchy and its nonnegative increments. The complete level-resolved distribution is obtained only at Krylov saturation. If the exact QFI is known independently, a truncated run yields a normalized prefix together with the aggregate unresolved tail mass; it does not determine the missing levelwise weights. Without the exact QFI, the run determines only the unnormalized resolved weights $\mathcal F_{k+1}-\mathcal F_k$ and a lower bound on their total mass.

The construction of the two orthonormal bases, the equivalence between state-metric projection and Hilbert--Schmidt Galerkin approximation, and the proof of Eq.~\eqref{eq:qfi-central-main} are given in Appendix~\ref{app:Krylov-review}.

\section{Spectral Representation and Convergence of the Krylov Approximation}
\label{sec:spectral}

The Krylov distribution introduced in Sec.~\ref{sec:QFI} resolves the QFI along successive operator-space directions. A complementary description is obtained by resolving the same response over the spectrum of $\mathcal K_\rho$. This spectral viewpoint makes clear which Liouville modes carry the metrological susceptibility, why two spectral measures appear, and why the convergence changes qualitatively when the active spectrum approaches zero. At any fixed finite dimension the active spectral measure is atomic and the Krylov sequence terminates after finitely many steps. All identities below are therefore exact at finite size. For a sequence of trace-one states, an overall spectral scale can shrink merely because the Hilbert-space dimension grows. Accordingly, ``gapped'' and ``hard-edge'' below refer to the rescaled active eigenvalues $\widehat\lambda=\lambda/\lambda_{\max}$, or equivalently to the behavior of the condition number and the shape of the normalized support. The ratios $\mathcal F_n/\mathcal F$ are invariant under this common positive rescaling.

Let $E_{\mathcal K_\rho}(\mathrm d\lambda)$ be the spectral resolution of $\mathcal K_\rho$ on the cyclic subspace generated by the tangent operator $B$---that is, the spectrum restricted to the subspace reachable from $B$ by repeated application of $\mathcal K_\rho$. Equivalently, let $\{\psi_j\}$ be Hilbert--Schmidt-orthonormal eigenoperators satisfying
\begin{equation}
 \mathcal K_\rho\psi_j=\lambda_j\psi_j,
 \qquad
 B=\sum_j c_j\psi_j,
 \label{eq:modal-decomposition}
\end{equation}
where only modes with $c_j\neq0$ belong to the active spectrum. The exact SLD is
\begin{equation}
 L=\sum_j\frac{c_j}{\lambda_j}\psi_j,
\end{equation}
and the contribution of the $j$th active mode to the QFI is
\begin{equation}
 \mathcal F_j=\frac{|c_j|^2}{\lambda_j},
 \qquad
 \mathcal F=\sum_j\mathcal F_j.
 \label{eq:modal-QFI}
\end{equation}
The eigenvalue $\lambda_j$ plays the role of a Liouville-space stiffness. A small value produces a large inverse response, but only when the physical tangent has nonzero overlap with that mode.

Two normalized spectral measures arise because the same source is normalized in two different operator geometries. They have identical support on the active positive subspace and do not represent different spectra. With
\begin{equation}
 u_0=\frac{B}{\beta_0}, \qquad \beta_0^2=(B,B)_0=\sum_j|c_j|^2,
\end{equation}
the Hilbert--Schmidt source measure is
\begin{equation}
 \mathrm d\nu(\lambda)
 =(u_0,E_{\mathcal K_\rho}(\mathrm d\lambda)u_0)_0
 =\sum_j\frac{|c_j|^2}{\beta_0^2}
 \delta(\lambda-\lambda_j)\,\mathrm d\lambda.
 \label{eq:nu-measure}
\end{equation}
The measure $\nu$ records how the raw tangent $B=\partial_\theta\rho$ is distributed over the active Liouville modes before the inverse response is applied. The QFI is its first inverse moment,
\begin{equation}
 \mathcal F
 =\beta_0^2\int\frac{\mathrm d\nu(\lambda)}{\lambda}.
 \label{eq:F-nu-exact}
\end{equation}
This is the measure naturally generated by the Hilbert--Schmidt Lanczos recursion and therefore the measure relevant to the computational hierarchy.

With
\begin{equation}
 v_0=\frac{B}{\beta_\rho}, \qquad \beta_\rho^2=\langle B,B\rangle_\rho=\sum_j\lambda_j|c_j|^2,
\end{equation}
the state-metric source measure is
\begin{equation}
 \mathrm d\mu(\lambda)
 =\langle v_0,E_{\mathcal K_\rho}(\mathrm d\lambda)v_0\rangle_\rho
 =\sum_j\frac{\lambda_j|c_j|^2}{\beta_\rho^2}
 \delta(\lambda-\lambda_j)\,\mathrm d\lambda.
 \label{eq:mu-measure}
\end{equation}
Equivalently, $\mu$ is the Hilbert--Schmidt spectral measure of the metric-dressed source $\mathcal K_\rho^{1/2}B/\beta_\rho$. It is therefore the natural measure for the state-orthonormal expansion of the SLD. In this representation,
\begin{equation}
 \mathcal F
 =\beta_\rho^2\int\frac{\mathrm d\mu(\lambda)}{\lambda^2}.
 \label{eq:F_mu_exact}
\end{equation}
The filtration-adapted state-orthonormal Krylov vectors and the corresponding SLD coefficients are generated by polynomials orthonormal with respect to $\mu$. Thus $\mu$ organizes the nested partial Parseval sums and the exact error tail of the QFI Krylov distribution.

Let
\begin{equation}
 m_1=\int\lambda\,\mathrm d\nu(\lambda)
 =\frac{\beta_\rho^2}{\beta_0^2}.
 \label{eq:m1}
\end{equation}
The relation between the two operator metrics implies the normalized Christoffel transformation
\begin{equation}
 \mathrm d\mu(\lambda)
 =\frac{\lambda}{m_1}\,\mathrm d\nu(\lambda).
 \label{eq:measure-relation-main}
\end{equation}
This relation immediately implies that the two inverse-moment representations are mathematically equivalent despite their different spectral weightings. Indeed,
\begin{equation}
 \beta_\rho^2\frac{\mathrm d\mu(\lambda)}{\lambda^2}
 =\beta_0^2\frac{\mathrm d\nu(\lambda)}{\lambda}.
 \label{eq:modewise-equivalence}
\end{equation}
The first- and second-inverse-moment formulas therefore agree mode by mode. The additional inverse power in Eq.~\eqref{eq:F_mu_exact} is exactly compensated by the extra factor of $\lambda$ in the state-metric measure.

The two measures nevertheless have different finite-order roles. The Hilbert--Schmidt measure generates the Lanczos matrix and the Gauss--Lanczos approximation to the first inverse moment. The state-metric measure generates the orthogonal expansion of $1/\lambda$ whose partial Parseval sums are the resolved QFI weights. Applying Gaussian quadrature directly to $\lambda^{-2}$ with respect to $\mu$ generally produces a different finite-dimensional approximation and must not be identified with the QFI projection hierarchy. The full orthogonal-polynomial details, including the Christoffel transformation and the relation between the two polynomial systems, are given in Appendix~\ref{app:spectral-rho}.

It is also useful to isolate the spectral distribution of the QFI itself:
\begin{equation}
 \mathrm d\pi_{\mathcal F}(\lambda)
 =\frac{\beta_0^2}{\mathcal F}
 \frac{\mathrm d\nu(\lambda)}{\lambda}
 =\frac{\beta_\rho^2}{\mathcal F}
 \frac{\mathrm d\mu(\lambda)}{\lambda^2},
 \qquad
 \int\mathrm d\pi_{\mathcal F}=1.
 \label{eq:qfi-contribution-measure}
\end{equation}
Unlike $\nu$ and $\mu$, which characterize the source in different operator geometries, $\pi_{\mathcal F}$ directly quantifies how the QFI itself is distributed over the active Liouville spectrum. It is therefore the true spectral distribution of metrological information.

Writing $\rho=\sum_a \varrho_a|a\rangle\langle a|$ and $E_{ab}=|a\rangle\langle b|$, the canonical eigenoperators diagonalize the SLD superoperator:
\begin{equation}
 \mathcal K_\rho E_{ab}=w_{ab}E_{ab},
 \qquad
 w_{ab}=\frac{\varrho_a+\varrho_b}{2}.
 \label{eq:pair-sums}
\end{equation}
The two measures therefore take the explicit forms given in Appendix~\ref{app:spectral-rho}, with the state-metric measure containing one factor of $w_{ab}$ reflecting one insertion of the metric. For unitary encoding, $B_{ab}=i(\varrho_a-\varrho_b)H_{ab}$. The active lower edge is controlled jointly by the pair-sum spectrum and the tangent overlap. Small eigenvalues of $\rho$ do not by themselves imply slow convergence: the relevant soft modes must also be excited by the parameter variation.

Every vector in the $n$th Krylov space can be written as $Q_{n-1}(\mathcal K_\rho)B$, where $q_{n-1}$ is a polynomial of degree at most $n-1$. The projection error therefore has the equivalent spectral forms
\begin{align}
 \mathcal F-\mathcal F_n
 &=\beta_0^2\min_{q\in\Pi_{n-1}}
 \int \lambda\left|\frac{1}{\lambda}-Q(\lambda)\right|^2
 \mathrm d\nu(\lambda)
 \nonumber\\
 &=\beta_\rho^2\min_{q\in\Pi_{n-1}}
 \int\left|\frac{1}{\lambda}-Q(\lambda)\right|^2
 \mathrm d\mu(\lambda).
 \label{eq:best-polynomial-error}
\end{align}
The convergence problem is thereby reduced to a weighted polynomial approximation problem for the singular function $1/\lambda$. The decisive issue is whether the pole at $\lambda=0$ remains separated from the active support or becomes part of its limiting edge.

\subsection*{Gapped spectrum: exponential convergence}

Assume that the active spectrum is contained in
\begin{equation}
 0<\lambda_{\min}\leq\lambda\leq\lambda_{\max}.
\end{equation}
All active Liouville directions then have a finite stiffness, and $1/\lambda$ is analytic in a neighborhood of the support. The standard conjugate-gradient energy-norm estimate gives~\cite{saad2003iterative}
\begin{equation}
 0\leq1-\frac{\mathcal F_n}{\mathcal F}
 \leq\min\{1,4q^{2n}\},
 \qquad
 q=\frac{\sqrt\kappa-1}{\sqrt\kappa+1},
 \quad
 \kappa=\frac{\lambda_{\max}}{\lambda_{\min}}.
 \label{eq:exp-error}
\end{equation}
This is the regular, exponentially convergent regime. The estimate becomes slow when the lower edge approaches zero and becomes exact after one step when the active spectrum collapses to a single point. We use Eq.~\eqref{eq:exp-error} as a known baseline rather than as a new result.

For every finite full-rank state, the smallest active eigenvalue is positive and the Krylov sequence ultimately saturates. A hard-edge law instead describes a sequence of systems for which $\lambda_{\min}$ tends to zero and a nontrivial limiting measure develops near the origin.

\subsection*{Hard edge at zero: exact Jacobi class}

To analyze the singular convergence regime in a controlled setting, we
consider an exactly solvable family of state-metric spectral measures whose
support reaches the singular point of the QFI resolvent at
$\lambda=0$. In contrast to the gapped case, where the inverse response
$1/\lambda$ is analytic on the spectral support and the Krylov hierarchy
converges exponentially, a spectral density extending to zero can produce
algebraic convergence because low-$\lambda$ Liouville modes are strongly
amplified by the inverse response.

We consider the normalized Jacobi family
\begin{equation}
 \mathrm d\mu_\alpha(\lambda)
 =(\alpha+1)\lambda^\alpha\,\mathrm d\lambda,
 \qquad
 0\leq\lambda\leq1,
 \qquad
 \alpha>1 .
 \label{eq:mu-alpha}
\end{equation}
The condition $\alpha>1$ ensures that the QFI remains finite, since in the state-metric representation the QFI is determined by the second inverse moment
\begin{equation}
\mathcal F=\beta_\rho^2\int\frac{\mathrm d\mu(\lambda)}{\lambda^2}.
\end{equation}
This continuum statement concerns the energy-metric completion of the cyclic space. The Hilbert--Schmidt norm of the limiting SLD instead contains $\int \mathrm d\nu(\lambda)/\lambda^2$, which is finite for this family only when $\alpha>2$. Thus for $1<\alpha\leq2$ the formulas describe the limit of finite-cutoff systems (or an element of the state-metric completion), not a Hilbert--Schmidt SLD in the unregularized continuum model.
The corresponding Hilbert--Schmidt source measure, related through the Christoffel transformation, is
\begin{equation}
 \mathrm d\nu_\alpha(\lambda)
 =\alpha\lambda^{\alpha-1}\,\mathrm d\lambda .
 \label{eq:nu-alpha}
\end{equation}
Therefore, the same hard-edge structure appears as a power-law accumulation of the computational spectral measure near zero.

It is also useful to introduce the normalized QFI contribution measure,
\begin{equation}
 \mathrm d\pi_{\mathcal F}(\lambda)
 =(\alpha-1)\lambda^{\alpha-2}\,\mathrm d\lambda .
 \label{eq:Jacobi-qfi-measure}
\end{equation}
This measure directly describes how the metrological susceptibility is distributed among the Liouville modes. For $1<\alpha<2$, the total QFI is finite and its density is enhanced, indeed divergent, toward the hard edge; nevertheless the mass of $[0,\delta]$ equals $\delta^{\alpha-1}$ and vanishes as $\delta\to0$. At $\alpha=2$, the QFI contribution is uniformly distributed over the spectrum, while for $\alpha>2$ its low-energy density is suppressed. Thus the exponent $\alpha$ controls both the convergence rate and the local enhancement of metrological information near weak Liouville modes.

For the Jacobi family, the orthogonal polynomials associated with $\mathrm d\mu_\alpha$ are shifted Jacobi polynomials, allowing an exact evaluation of the Krylov coefficients. The resulting finite-order QFI error is
\begin{equation}
 1-\frac{\mathcal F_n}{\mathcal F}
 =
 \left[
 \frac{\Gamma(\alpha)\Gamma(n+1)}
 {\Gamma(n+\alpha)}
 \right]^2 .
 \label{eq:exact-Jacobi-error}
\end{equation}
This expression gives the complete convergence curve at every finite Krylov order and not only its asymptotic behavior.

For large Krylov dimension, the gamma-ratio asymptotics yield
\begin{equation}
 |\ell_k|\sim C_\alpha k^{1/2-\alpha},
 \qquad
 1-\frac{\mathcal F_n}{\mathcal F}
 \sim
 \Gamma(\alpha)^2 n^{-2(\alpha-1)} .
 \label{eq:Jacobi-asymptotic}
\end{equation}
Thus the hard-edge singularity changes the exponential convergence of the gapped case into an algebraic decay. For this Jacobi family, the Hilbert--Schmidt spectral density obeys
\begin{equation}
\frac{\mathrm d\nu}{\mathrm d\lambda}\sim\lambda^\gamma ,
\end{equation}
the exponent is $\gamma=\alpha-1$, and the convergence law becomes
\begin{equation}
1-\frac{\mathcal F_n}{\mathcal F}
\sim n^{-2\gamma}.
\end{equation}
Equivalently, the exact Jacobi result has the prefactor $\Gamma(\gamma+1)^2$. The exponent is not implied by the local power law alone; extension to a general measure requires the regularity and global assumptions stated below.

For the special case $\alpha=2$, the exact result reduces to the simple algebraic law
\begin{equation}
1-\frac{\mathcal F_n}{\mathcal F}
=
\frac{1}{(n+1)^2}.
\end{equation}
This provides a particularly transparent benchmark: although the QFI is finite, resolving the last fraction of the metrological information requires progressively deeper Krylov levels because the inverse response amplifies the contribution of low-$\lambda$ modes.

The value $\alpha=1$ marks the boundary between the finite-QFI
hard-edge regime and the divergent regime. It is not included in the
Jacobi convergence theorem above because the QFI inverse moment becomes
logarithmically divergent. For

\begin{equation}
\mathrm d\mu_1(\lambda)
=
2\lambda\,\mathrm d\lambda ,
\end{equation}
one obtains
\begin{equation}
\mathcal F
=2 \beta_\rho^2\int_0^1
\frac{\mathrm d\lambda}{\lambda},
\end{equation}
which diverges at the hard edge. In a finite system, an infrared cutoff
$\lambda_{\min}$ regularizes this divergence and gives the characteristic
behavior

\begin{equation}
\mathcal F
\sim
2\beta_\rho^2
\ln\frac{1}{\lambda_{\min}} .
\end{equation}

The marginal case therefore represents the transition between finite and
divergent QFI. Unlike the $\alpha>1$ Jacobi family, its finite-order Krylov
convergence depends on the specific infrared regularization and does not
possess the cutoff-independent closed form derived above.

The exact expression above is a theorem for the Jacobi family and should not be interpreted as a universal consequence of the local edge behavior alone. For a more general measure of the form
\begin{equation}
\mathrm d\mu(\lambda)
=
\lambda^\alpha h(\lambda)\mathrm d\lambda ,
\end{equation}
the asymptotic convergence depends additionally on the regularity, positivity, and global structure of the function $h(\lambda)$. The Jacobi result therefore provides an analytically controlled hard-edge benchmark against which more general spectral densities can be compared.

The exact tail also reveals that finite QFI does not necessarily imply a finite information-resolution depth. From Eq.~\eqref{eq:D-area},
\begin{equation}
 \mathcal D_\alpha
 =
 \sum_{n=1}^{\infty}
 \left[
 \frac{\Gamma(\alpha)\Gamma(n+1)}
 {\Gamma(n+\alpha)}
 \right]^2 .
 \label{eq:Jacobi-mean-depth}
\end{equation}
Using the asymptotic behavior of the summand, this quantity converges only for
\begin{equation}
\alpha>\frac32 .
\end{equation}
Therefore, in the regime
\begin{equation}
1<\alpha\leq\frac32 ,
\end{equation}
the total QFI is finite, but the mean Krylov depth diverges. Physically, this means that the total amount of metrological information remains bounded, while resolving it requires contributions from increasingly deep Krylov directions.

For the representative case $\alpha=2$, the mean depth is finite and evaluates to
\begin{equation}
 \mathcal D_2=\frac{\pi^2}{6}-1 .
\end{equation}

\section{Numerical Illustration of the Dual Krylov Framework}
\label{sec:applications}

The preceding sections establish the dual Krylov formulation analytically. Here we illustrate its finite-size numerical implementation for a mixed-field Ising generator and full-rank random states. The numerical calculation is not used as evidence for the analytic identities; rather, it demonstrates the matrix-free construction, the concentration of the resolved QFI weight, and the distinction between the bulk of the Krylov distribution and its extreme tail. In particular, the numerical calculation is continued to the point at which the unresolved QFI fraction reaches the prescribed operational convergence threshold, thereby directly revealing the Krylov order required to resolve the finite-size problem to that accuracy.

We consider the open-chain Hamiltonian
\begin{equation}
H=
-J\sum_{i=1}^{N-1}\sigma_i^z\sigma_{i+1}^z
-\sum_{i=1}^{N}\left(g\sigma_i^x+h\sigma_i^z\right),
\label{eq:Ising}
\end{equation}
with~\cite{Banuls:2010zki}
\begin{equation}
J=1,\qquad g=-1.05,\qquad h=0.5.
\end{equation}
For each realization, a full-rank state is sampled from the complex Ginibre induced ensemble,
\begin{equation}
\rho=\frac{XX^\dagger}{\Tr(XX^\dagger)},
\label{eq:Wishart-state}
\end{equation}
where the real and imaginary parts of the entries of $X$ are independent normal random variables with equal variance. The unitary tangent is
\begin{equation}
B=i[\rho,H].
\label{eq:numerical-seed}
\end{equation}

We use $N=12$, for which $d=2^{12}=4096$
and the number of unordered active spectral pairs is bounded by
\begin{equation}
M=\frac{d(d-1)}{2}=8,386,560.
\label{eq:pair-count-num}
\end{equation}
The actual cyclic-space saturation dimension satisfies $d_0\leq M$. Importantly, QFI convergence need not require resolving the full cyclic dimension: the relevant numerical question is how rapidly the QFI weight is accumulated along the Krylov hierarchy.

Writing
\[
\rho=V\operatorname{diag}(\varrho_1,\ldots,\varrho_d)V^\dagger
\]
and
\[
\widetilde H=V^\dagger H V,
\]
the compressed kernel nodes and weights are
\begin{equation}
w_{ab}=\frac{\varrho_a+\varrho_b}{2},
\qquad
\Omega_{ab}=2(\varrho_a-\varrho_b)^2|\widetilde H_{ab}|^2,
\qquad a<b,
\label{eq:Omega-ab-num}
\end{equation}
where $\varrho_a$ are the eigenvalues of $\rho$. Thus
\begin{equation}
\beta_0^2=\sum_{a<b}\Omega_{ab},
\qquad
\mathcal F=\sum_{a<b}\frac{\Omega_{ab}}{w_{ab}}.
\label{eq:exactQFI-num}
\end{equation}
The full Liouville-space superoperator is therefore never constructed. Instead, Lanczos is applied directly to the diagonal node operator, with normalized seed components $\sqrt{\Omega_{ab}}/\beta_0$.

The finite-order hierarchy is
\begin{equation}
\mathcal F_n
=\beta_0^2 e_0^\dagger J_n^{-1}e_0,
\label{eq:Fn-num}
\end{equation}
and its resolved increments are
\begin{equation}
p_n=\frac{\mathcal F_{n+1}-\mathcal F_n}{\mathcal F}.
\label{eq:pn-num}
\end{equation}
Consequently,
\begin{equation}
\sum_{n=0}^{m-1}p_n
=\frac{\mathcal F_m}{\mathcal F}
=1-\epsilon_m,
\qquad
\epsilon_m=1-\frac{\mathcal F_m}{\mathcal F}.
\label{eq:error-num}
\end{equation}
The sequence $\epsilon_m$ is the unresolved QFI weight after $m$ Krylov steps.

\begin{figure}[t]
\centering
\includegraphics[width=0.42\textwidth]{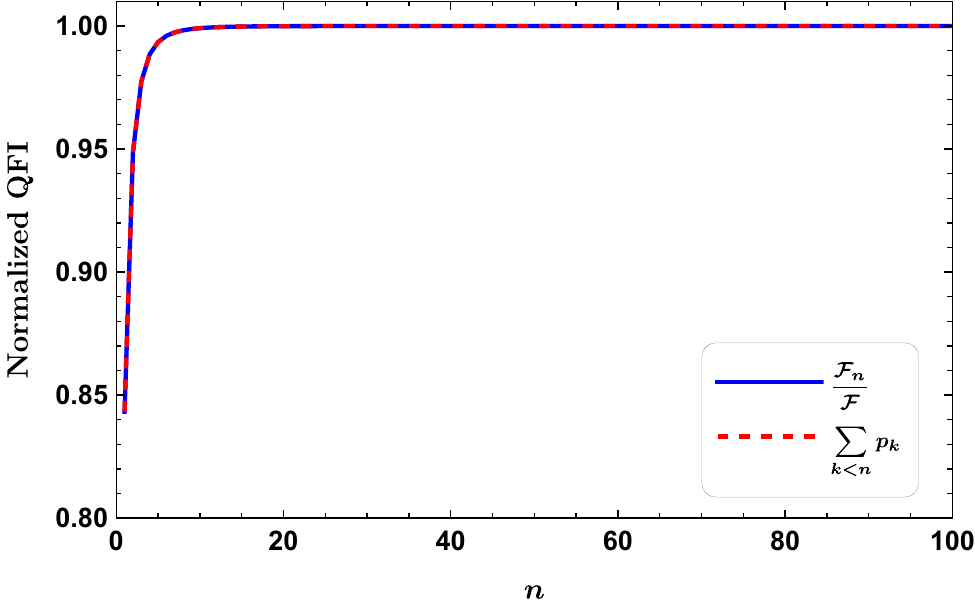}
\caption{Telescoping consistency visualization for $N=12$, averaged over ten independent Ginibre realizations at $J=1$, $g=-1.05$, and $h=0.5$. The normalized hierarchy $\mathcal F_n/\mathcal F$ (solid) is compared with $\sum_{k<n}p_k$ (dashed). Their agreement checks the indexing and accumulation of the hierarchy increments; it is not an independent verification of the dual Krylov identity.}
\label{fig:identity}
\end{figure}

Figure~\ref{fig:identity} compares $\mathcal F_n/\mathcal F$ with the cumulative sum of the increments. Their agreement is a telescoping consistency check of the indexing and accumulation. A noncircular verification of the filtration-resolved identity would additionally require a separately constructed state-orthonormal basis and direct evaluation of the corresponding SLD overlaps.

\begin{figure}[t]
\centering
\includegraphics[width=0.42\textwidth]{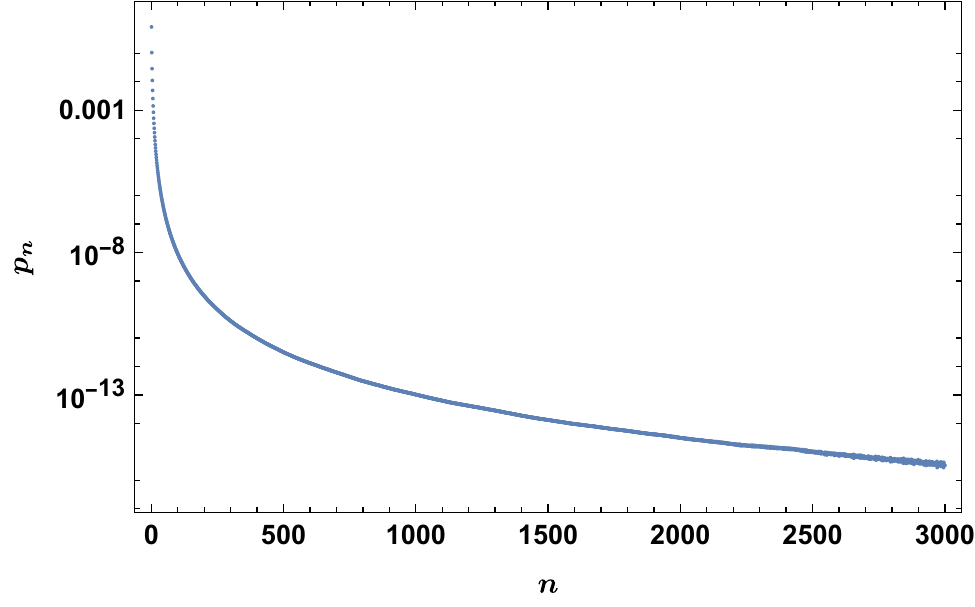}
\caption{Ensemble mean of the resolved QFI weights $p_k$ for $N=12$, averaged over ten independent Ginibre realizations. The weights are normalized by the independently evaluated spectral QFI and are not renormalized after truncation. The first three mean weights account for approximately $97.6\%$ of the QFI, demonstrating the strong concentration of metrological weight at low Krylov order. The plotted range is shown for visual clarity; the convergence analysis uses the continued hierarchy.}
\label{fig:distribution}
\end{figure}

The ensemble-mean first weights are
\begin{equation}
\overline p_0\simeq0.843,
\qquad
\overline p_1\simeq0.105,
\qquad
\overline p_2\simeq0.028.
\label{eq:first-weights-num}
\end{equation}
Their sum is approximately $0.976$, showing that about $97.6\%$ of the QFI weight is already resolved in the first three Krylov directions. Figure~\ref{fig:distribution} displays the resulting QFI-weight profile over the low-to-intermediate Krylov orders. The plotted range is chosen for visualization; the underlying calculation is continued beyond the displayed range when determining the convergence of the full hierarchy.

The information depth resolved through order $m$ is
\begin{equation}
\mathcal D_m=\sum_{n=0}^{m-1}n\,p_n.
\label{eq:info-depth-num}
\end{equation}
For the continued $N=12$ calculation, the resolved information depth reaches
\begin{equation}
\mathcal{ D}_{4500}\simeq0.263.
\label{eq:depth-value-num}
\end{equation}
This value should be interpreted as the information depth accumulated by the resolved Krylov hierarchy at the stated convergence order. It is already well below unity, reflecting the strong concentration of QFI weight at very low Krylov order, while the unresolved QFI weight is below the operational threshold and, at this numerical resolution, constitutes an extremely small residual tail.

\begin{figure}[t]
\centering
\includegraphics[width=0.42\textwidth]{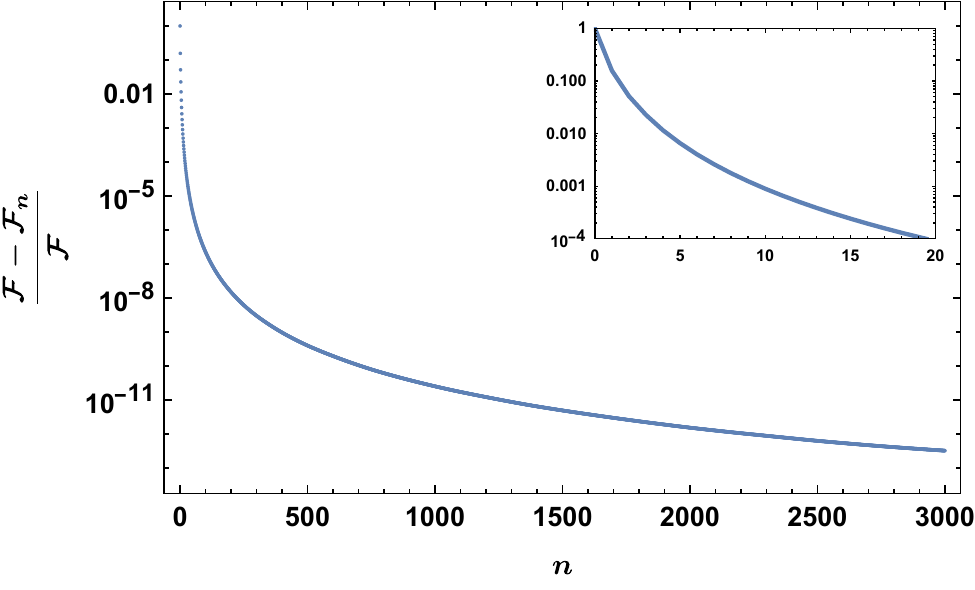}
\caption{Ensemble-mean unresolved QFI fraction 
$\epsilon_n=1-\mathcal F_n/\mathcal F$ for $N=12$,
averaged over ten independent Ginibre realizations.
The main panel emphasizes the rapid initial capture of
QFI weight and the subsequent weak tail, while the 
inset resolves the first $20$ Krylov orders. 
The hierarchy was continued until the operational 
convergence criterion $\epsilon_n<10^{-15}$ was 
reached; the observed crossing occurs 
at $n_{\mathrm{conv}}^{\mathrm{(op)}}\simeq4500$.}
\label{fig:error}
\end{figure}

The unresolved QFI fraction provides a complementary measure of convergence. Figure~\ref{fig:error} shows its rapid initial decrease and the subsequent weak tail. Continuing the same numerical calculation until the prescribed operational threshold is reached gives
\begin{equation}
n_{\mathrm{conv}}^{\mathrm{(op)}}\simeq 4500,
\qquad
\epsilon_{4500}<10^{-15}.
\label{eq:nconv-num}
\end{equation}
Thus the numerical calculation itself identifies $n\simeq4500$ as the Krylov order at which the unresolved QFI fraction first falls below the chosen near-machine-precision threshold. This is not an imposed truncation of the calculation, but an observed operational convergence scale for the finite-size realization. Relative to the full unordered-pair count,
\begin{equation}
\frac{n_{\mathrm{conv}}^{\mathrm{(op)}}}{M}
\simeq 5.3\times10^{-4},
\end{equation}
so the hierarchy reaches the stated numerical accuracy at only about $0.10\%$ of the maximum pair-space dimension.

The two quantities
\begin{equation}
\mathcal{ D}_{4500}\simeq0.263
\qquad\text{and}\qquad
n_{\mathrm{conv}}^{\mathrm{(op)}}\simeq4500
\end{equation}
characterize different aspects of the same QFI distribution. The information depth measures where the dominant metrological weight is concentrated, whereas the operational convergence order measures how deeply one must resolve the hierarchy to capture the extremely small residual tail. Their large separation is therefore not contradictory: a distribution can have a very small mean depth while still possessing a long, numerically tiny tail.

At machine precision, a residual below $10^{-15}$ is only a few units in the last place. We therefore interpret Eq.~\eqref{eq:nconv-num} as a computed residual crossing rather than as a certification of fifteen-digit accuracy in the QFI itself. A genuine high-precision verification would require regenerating the full calculation at higher working precision.

The numerical results therefore reveal two distinct
features of the dual Krylov hierarchy. First, the 
metrological weight is overwhelmingly concentrated at 
low Krylov order, with the first three directions
already carrying approximately $97.6\%$ of the QFI 
and the resolved information depth remaining small. 
Second, the hierarchy possesses a much longer 
but extremely weak tail: the calculation reaches the
operational threshold $\epsilon_n<10^{-15}$ only at
$n\simeq 4500$. The numerical calculation thus directly 
demonstrates the separation between the scale governing 
the bulk of the QFI distribution and the much larger 
scale required to resolve its extreme tail. These 
observations are complementary to, and do not replace,
the analytic dual identity and the Jacobi hard-edge 
convergence analysis.

Finally, for every finite full-rank realization considered here,
 one has $w_{ab}>0$,
so the compressed spectrum has a strictly positive minimum. The numerical examples therefore represent finite-size realizations of the gapped regime. The hard-edge regime studied analytically requires a sequence of spectra in which spectral weight develops arbitrarily close to $\lambda=0$. The numerical observation of a very small but nonzero unresolved QFI fraction should therefore not be interpreted as a hard-edge calculation; rather, it illustrates how the finite-size gapped hierarchy can nevertheless exhibit a large separation between the dominant QFI scale and the operational convergence scale.
\section{Conclusions}
\label{sec:conclusions}

Building on the established Krylov hierarchy of QFI lower bounds, this work
develops its filtration-resolved operator-geometric interpretation. The
algebraic Krylov space associated with the SLD equation admits two
complementary ordered orthonormal representations: the Hilbert--Schmidt
geometry governing the computational Lanczos construction and the
state-dependent geometry governing the metrological content of the SLD. The
level-resolved information distribution, its information depth, the dual
Christoffel-related spectral measures, and the exact Jacobi hard-edge
benchmark are the principal contributions developed here.

The main level-resolved identity is

\begin{equation}
 \mathcal F_n
 =
 \beta_0^2 e_0^\dagger J_n^{-1}e_0
 =
 \sum_{k=0}^{n-1}|\ell_k|^2 .
\end{equation}

This relation establishes an explicit bridge between the computational QFI
hierarchy generated by Hilbert--Schmidt Lanczos and the expansion of the exact
SLD in the filtration-adapted state-orthonormal Krylov basis. As a
consequence, the successive
increments of the hierarchy define a genuine probability distribution,

\begin{equation}
p_k=
\frac{|\ell_k|^2}{\mathcal F}
=
\frac{\mathcal F_{k+1}-\mathcal F_k}{\mathcal F},
\end{equation}
which quantifies how quantum Fisher information is resolved across Krylov
directions. The associated information depth,
\begin{equation}
\mathcal D=\sum_k k p_k ,
\end{equation}
provides an intrinsic measure of the average Krylov order at which
metrological information resides. A single Hilbert--Schmidt Lanczos recursion
provides every completed increment; no separate state-metric Lanczos solve is
required. The complete level-resolved profile is obtained only at saturation.
If the exact $\mathcal F$ is known independently, a truncated run yields a
normalized prefix plus the total unresolved tail mass, but not the missing
levelwise profile.

The same duality appears naturally in the spectral formulation. The
Hilbert--Schmidt and state-dependent descriptions generate two
Christoffel-related measures on the active spectrum of the QFI kernel. The
Hilbert--Schmidt measure governs the Lanczos quadrature of the first inverse
spectral moment, while the state-dependent measure governs the SLD expansion
and the second inverse moment representation of the QFI. Their relation,

\begin{equation}
 \beta_\rho^2
 \frac{\mathrm d\mu(\lambda)}{\lambda^2}
 =
 \beta_0^2
 \frac{\mathrm d\nu(\lambda)}{\lambda},
\end{equation}
resolves the apparent difference between the two inverse powers of
$\lambda$. The two formulations are therefore not competing descriptions but
different spectral representations of the same physical information content.

Within this spectral framework, Krylov convergence becomes a problem of
weighted polynomial approximation to the singular function $1/\lambda$. In
the gapped regime, the conventional condition-number picture is recovered. The
principal new convergence result of this work concerns the complementary
singular regime, where the spectral support reaches the origin. For the
exactly solvable Jacobi hard-edge family,

\begin{equation}
d\mu_\alpha=(\alpha+1)\lambda^\alpha d\lambda ,
\qquad \alpha>1 ,
\end{equation}
we obtain the exact finite-order convergence law
\begin{equation}
1-\frac{\mathcal F_n}{\mathcal F}
=
\left[
\frac{\Gamma(\alpha)\Gamma(n+1)}
{\Gamma(n+\alpha)}
\right]^2
\sim
\Gamma(\alpha)^2 n^{-2(\alpha-1)} .
\end{equation}

This result provides a rigorous benchmark for algebraic convergence caused by
spectral accumulation near the hard edge. The marginal case $\alpha=1$
marks the transition where the QFI becomes logarithmically divergent and the
finite-size convergence behavior changes qualitatively. These results show
that the convergence rate of Krylov QFI approximations is controlled not only
by the Krylov dimension but by the spectral 
geometry of the information
measure near small kernel eigenvalues.

The finite-size numerical study demonstrates how these abstract structures appear in interacting quantum systems. Using a matrix-free implementation for a mixed-field Ising model with $N=12$ spins, the QFI hierarchy, Krylov distribution, and finite-order error are obtained directly from the active spectral pairs of the QFI kernel without constructing the full Liouville-space superoperator. Ensemble-averaged calculations over random Ginibre states verify the central identity to machine precision and show that the metrological information is concentrated in a comparatively small Krylov sector. For the representative realization, the compressed spectral space comprises $M=8\,386\,560$ active pairs, while a relative QFI resolution of $10^{-15}$ is achieved at the operational convergence order $n_{\mathrm{conv}}^{(\mathrm{op})}\simeq4500$, corresponding to approximately $0.10\%$ of the maximum pair-space dimension. The resolved information depth at this order is $\mathcal D_{4500}\simeq0.263$, indicating that the QFI weight is overwhelmingly concentrated in the first few Krylov directions. For every finite realization, full rank of $\rho$ gives a strictly positive minimum of the compressed spectrum, so the observed convergence is consistent with the finite-size realization of the gapped regime; the hard-edge algebraic regime would require a sequence of spectra developing an accumulation of eigenvalues toward zero. The large separation between the available spectral dimension and the low-order concentration of the information distribution nevertheless highlights the strong intrinsic compressibility of the QFI hierarchy in Krylov space.

In conclusion, the dual Krylov framework is more than a numerical
acceleration scheme for operator inversion. It provides a geometric
decomposition of where quantum information resides in operator space and how
it is progressively resolved. The Krylov distribution $p_k$ locates the
metrological sensitivity in the ordered Krylov filtration, the information
depth $\mathcal D$ quantifies the associated resolution scale, the dual
spectral measures identify the Liouville modes responsible for the response,
and the convergence theory determines how these contributions accumulate as
the Krylov space grows.

The operator-geometric perspective developed here is not restricted to QFI.
Because the construction relies on the geometry of Hermitian linear-response
problems, analogous dual representations may arise for other quantum
information quantities defined through operator inverse problems. Future
directions include extensions to continuous spectra, infinite-dimensional
systems, open quantum dynamics, and quantum critical regimes, where the
accumulation of small kernel eigenvalues may generate singular information
geometries and enhanced metrological responses.

\textbf{Note added:}
An independent study appearing shortly before the first version of the
present work established condition-number bounds for spectra separated from
zero, analyzed exact recovery in low-rank settings, and developed practical
convergence considerations for Krylov QFI calculations~\cite{Wang:2026ysd}.
Those results concern the performance and certification of the established
Krylov bounds. The present work addresses a complementary question: their
filtration-resolved information geometry. Specifically, we introduce the
Krylov distribution $p_k$ and information-resolution depth $\mathcal D$,
establish the dual spectral measures connecting the computational and
state-dependent geometries, and derive an exact hard-edge convergence law for
the Jacobi family. We thank the authors for bringing their work to our
attention.

\section*{Acknowledgements}

We would like to thank Souvik Banerjee and  Mohammad Reza Tanhayi for 
many insightful discussions on various aspects of  
Krylov space and Krylov complexity. The work of 
M. A.  is supported by the Iran National Science Foundation (INSF) under 
Project No.~4023620. The work of M. J. V. is based on research funded by the Iran National Science Foundation (INSF) under Project No.~4047915.  We also
acknowledge the assistance of ChatGPT for help to improve the quality of the written text.

\appendix


\section{Dual Krylov Geometry of the SLD Equation}
\label{app:Krylov-review}

The SLD equation is a linear system in operator space. Its mathematical structure is therefore the same as solving $Ax=b$, but the metric in which the solution is evaluated is not the metric in which the Lanczos algorithm runs. This appendix explains why these two metrics must be kept distinct and how they nevertheless lead to a unified finite-order identity.

Consider a Hermitian positive-definite operator $A$ acting on a finite-dimensional space with product $(\cdot,\cdot)_0$, and let
\begin{equation}
Ax=b,
\label{eq:app-linear-system}
\end{equation}
where $A \in \mathbb{C}^{N \times N}$ represents a physical operator and $b \in \mathbb{C}^N$ is a source term.  
Such linear systems arise in discretizations of differential operators, Green's-function evaluations, response theory, and quantum many-body dynamics~\cite{saad2003iterative,dk94,dk98,guttel2017}.

The Krylov spaces
\begin{equation}
\mathcal K_n(A,b)=\operatorname{span}\{b,Ab,\ldots,A^{n-1}b\}
\label{eq:app-general-Krylov}
\end{equation}
capture the directions reachable from the source $b$ by repeated action of $A$.

The Lanczos algorithm constructs an orthonormal basis of these nested spaces. Starting from the normalized initial vector
\begin{equation}
q_0=\frac{b}{\beta}, \qquad \beta=\|b\|_0,
\label{eq:app-q0}
\end{equation}
the algorithm generates the sequence via the three-term recurrence
\begin{equation}
A q_k = b_k q_{k-1} + a_k q_k + b_{k+1} q_{k+1},
\label{eq:app-lanczos-recurrence}
\end{equation}
where the Lanczos coefficients are given by
\begin{equation}
a_k=(q_k,Aq_k)_0, \qquad b_{k+1}=\|Aq_k-a_kq_k-b_kq_{k-1}\|_0,
\label{eq:app-lanczos-coeffs}
\end{equation}
with $b_0=0$ and $q_{-1}=0$. At each step, $q_{k+1}$ is obtained by normalizing the residual
\begin{equation}
r_k=Aq_k-a_kq_k-b_kq_{k-1},
\label{eq:app-residual}
\end{equation}
so that $b_{k+1}=\|r_k\|_0$ and $q_{k+1}=r_k/b_{k+1}$. The procedure terminates at the Krylov dimension $d_0$, when $b_{d_0}=0$. The orthonormal basis $Q_n=(q_0,q_1,\ldots,q_{n-1})$ therefore satisfies $Q_n^\dagger Q_n=I_n$, and $A$ is represented in this basis by the real symmetric tridiagonal matrix
\begin{equation}
T_n=Q_n^\dagger A Q_n
=
\begin{pmatrix}
a_0 & b_1 & & & \\
b_1 & a_1 & b_2 & & \\
& b_2 & a_2 & \ddots & \\
& & \ddots & \ddots & b_{n-1} \\
& & & b_{n-1} & a_{n-1}
\end{pmatrix}.
\label{eq:app-Tn}
\end{equation}
The finite Lanczos relation is
\begin{equation}
A Q_n = Q_n T_n + b_n q_n e_{n-1}^{\mathsf T},
\label{eq:app-lanczos-relation}
\end{equation}
where the residual term measures the leakage from the $n$-dimensional subspace into the next Krylov direction.

The Galerkin approximation $x_n\in\mathcal K_n(A,b)$ is obtained by requiring the residual to be orthogonal to the Krylov space:
\begin{equation}
(y,b-Ax_n)_0=0, \qquad y\in\mathcal K_n(A,b).
\label{eq:app-Galerkin}
\end{equation}
Writing $x_n=Q_nz_n$, this condition becomes
\begin{equation}
Q_n^\dagger(b-AQ_nz_n)=0.
\end{equation}
Since $Q_n^\dagger b=\beta e_0$ and $Q_n^\dagger A Q_n=T_n$, we obtain
\begin{equation}
T_n z_n = \beta e_0,
\end{equation}
so
\begin{equation}
x_n=\beta Q_nT_n^{-1}e_0,
\label{eq:app-general-xn}
\end{equation}
where $e_0=(1,0,\ldots,0)^{\mathsf T}$.

The geometry naturally associated with the linear system is the energy metric
\begin{equation}
\|x\|_A^2=(x,Ax)_0.
\label{eq:app-energy-product}
\end{equation}
Since $Ax=b$, the Galerkin condition is equivalent to stating that $x_n$ is the orthogonal projection of the exact solution $x$ onto the Krylov space in this energy metric. Two norms of $x_n$ are therefore distinct: the ordinary Hilbert--Schmidt norm,
\begin{equation}
\|x_n\|_0^2=\beta^2e_0^\dagger T_n^{-2}e_0,
\label{eq:app-ordinary-norm}
\end{equation}
and the energy norm,
\begin{equation}
\|x_n\|_A^2=\beta^2e_0^\dagger T_n^{-1}e_0.
\label{eq:app-energy-norm}
\end{equation}
The energy norm is the physically relevant one when the linear system arises from a response problem, because it measures the solution in the metric induced by the operator $A$ itself.

For the SLD equation, we identify
\begin{equation}
A=\mathcal K_\rho, \qquad b=B=\partial_\theta\rho, \qquad x=L,
\label{eq:app-QFI-identification}
\end{equation}
where $\mathcal K_\rho(X)=\{\rho,X\}/2$; for unitary encoding, $B=i[\rho,H]$. The ordinary metric is the Hilbert--Schmidt product $(X,Y)_0=\Tr(X^\dagger Y)$, while the state-dependent product is
\begin{equation}
\langle X,Y\rangle_\rho=\frac12\Tr[\rho(X^\dagger Y+YX^\dagger)].
\label{eq:app-state-product}
\end{equation}
These two metrics are related by
\begin{equation}
\langle X,Y\rangle_\rho=(X,\mathcal K_\rho Y)_0.
\label{eq:app-metric-intertwining}
\end{equation}
Thus the state-dependent norm is precisely the $\mathcal K_\rho$-energy norm:
\begin{equation}
\|X\|_\rho^2=(X,\mathcal K_\rho X)_0.
\label{eq:app-state-energy}
\end{equation}
The QFI is simply this energy norm of the SLD:
\begin{equation}
\mathcal F=\|L\|_\rho^2=(B,\mathcal K_\rho^{-1}B)_0=(B,L)_0.
\label{eq:app-QFI-forms}
\end{equation}

The algebraic Krylov space
\begin{equation}
\Kry_n=\operatorname{span}\{B,\mathcal K_\rho B,\ldots,\mathcal K_\rho^{n-1}B\}
\label{eq:app-QFI-Krylov}
\end{equation}
is independent of the choice of inner product. It is the set of operator-space directions reachable from the physical tangent $B$ by repeated application of the superoperator $\mathcal K_\rho$. At saturation, the exact SLD belongs to this space, since one can always find a polynomial $p$ such that
\begin{equation}
L=\mathcal K_\rho^{-1}B=p(\mathcal K_\rho)B\in\Kry_{d_0}.
\label{eq:app-L-cyclic}
\end{equation}
This justifies expanding the SLD in a Krylov basis.

We first construct the computational basis in the Hilbert--Schmidt geometry. Let
\begin{align}
&U_n=(u_0,u_1,\ldots,u_{n-1}), \qquad U_n^\dagger U_n=I_n,\cr
& u_0=B/\beta_0, \qquad \beta_0=\|B\|_0,
\label{eq:app-Un}
\end{align}
be the Hilbert--Schmidt Lanczos basis of $\Kry_n$. Its tridiagonal representation is
\begin{equation}
J_n=U_n^\dagger\mathcal K_\rho U_n.
\label{eq:app-Jn}
\end{equation}
The Hilbert--Schmidt Galerkin solution is
\begin{equation}
\widehat L_n=\beta_0U_nJ_n^{-1}e_0.
\label{eq:app-Lhat}
\end{equation}
Its QFI norm is
\begin{equation}
\|\widehat L_n\|_\rho^2=\beta_0^2e_0^\dagger J_n^{-1}e_0,
\label{eq:app-Fn-HS}
\end{equation}
while its Hilbert--Schmidt squared norm would be
\begin{equation}
\|\widehat L_n\|_0^2=\beta_0^2e_0^\dagger J_n^{-2}e_0.
\label{eq:app-Ln-HS-norm}
\end{equation}
The difference between these two expressions is not a minor algebraic detail: the first is the physical QFI, the second is merely the Hilbert--Schmidt size of the approximation. Using the latter would correspond to measuring the response in the wrong metric.

The key physical result is that $\widehat L_n$ is not just any approximation: it is the state-metric orthogonal projection of the exact SLD. Indeed, for any $Y=U_ny\in\Kry_n$, using Eqs.~\eqref{eq:app-metric-intertwining} and $\mathcal K_\rho L=B$,
\begin{equation}
\langle Y,L-\widehat L_n\rangle_\rho
=\beta_0y^\dagger(e_0-J_nJ_n^{-1}e_0)=0.
\label{eq:app-projection-proof}
\end{equation}
Thus $\widehat L_n$ captures the part of the exact SLD that lies in the first $n$ Krylov directions, measured in the information geometry defined by $\rho$.

Now consider the information-geometric basis. Choose one state-orthonormal sequence adapted to the nested Krylov filtration,
\begin{equation}
 \operatorname{span}\{v_0,\ldots,v_k\}=\Kry_{k+1},
 \qquad
 v_k\in\Kry_{k+1}\cap\Kry_k^{\perp_\rho},
 \label{eq:app-filtration-adapted}
\end{equation}
with $\Kry_0=\{0\}$ and an arbitrary fixed phase convention. Before saturation, the new orthogonal complement is one-dimensional, so each level is intrinsic up to that phase. Let $V_n=(v_0,v_1,\ldots,v_{n-1})$ denote the first $n$ vectors of this single nested sequence. Then
\begin{equation}
V_n^\dagger\mathcal K_\rho V_n=I_n.
\label{eq:app-Vn-state-orthogonal}
\end{equation}
Expanding the exact SLD as
\begin{equation}
L=\sum_{k=0}^{d_0-1}\ell_kv_k, \qquad \ell_k=\langle v_k,L\rangle_\rho,
\label{eq:app-L-state-expansion}
\end{equation}
its state-metric projection is
\begin{equation}
L_n=\sum_{k=0}^{n-1}\ell_kv_k.
\label{eq:app-Ln-state}
\end{equation}
Uniqueness of the projection gives
\begin{equation}
L_n=\sum_{k=0}^{n-1}\ell_kv_k=\beta_0U_nJ_n^{-1}e_0.
\label{eq:app-vector-identity}
\end{equation}
Moreover, using $\mathcal K_\rho L=B$,
\begin{equation}
\ell_k=\langle v_k,L\rangle_\rho=(v_k,B)_0.
\label{eq:app-ell-seed}
\end{equation}
Thus the information-geometric coefficients are simply overlaps of the Krylov basis vectors with the physical tangent operator. Taking the state norm of Eq.~\eqref{eq:app-vector-identity} yields the central identity:
\begin{equation}
\sum_{k=0}^{n-1}|\ell_k|^2=\beta_0^2e_0^\dagger J_n^{-1}e_0.
\label{eq:app-central-identity}
\end{equation}
This tells us that the cumulative QFI captured by the first $n$ Krylov directions, computed from the information-geometric expansion, is exactly the standard computational bound obtained from the Hilbert--Schmidt Lanczos matrix. The resolved level weights are therefore obtained from successive bounds:
\begin{equation}
|\ell_k|^2=\mathcal F_{k+1}-\mathcal F_k.
\label{eq:app-increment}
\end{equation}
No second Lanczos recursion is needed. The complete level-resolved distribution is obtained at saturation. If $\mathcal F$ is supplied independently before saturation, the computed prefix can be normalized and the total unresolved remainder quantified, but its missing levelwise weights remain unknown.

Because both sequences are adapted to the same filtration, the two bases are related by an invertible upper-triangular matrix $C_n$ such that $V_n=U_nC_n$. A generic state-orthonormal rotation of $\Kry_n$ would not be triangular and would not preserve the individual level weights. State orthonormality gives
\begin{equation}
I_n=V_n^\dagger\mathcal K_\rho V_n=C_n^\dagger J_nC_n,
\label{eq:app-Crelation}
\end{equation}
so
\begin{equation}
J_n^{-1}=C_nC_n^\dagger.
\label{eq:app-Jinv-C}
\end{equation}
The coefficient vector is
\begin{equation}
\boldsymbol\ell^{(n)}=V_n^\dagger B=\beta_0C_n^\dagger e_0.
\label{eq:app-ellC}
\end{equation}
With the Cholesky factorization $J_n=R_n^\dagger R_n$, a convenient choice is
\begin{equation}
V_n=U_nR_n^{-1}, \qquad \boldsymbol\ell^{(n)}=\beta_0R_n^{-\dagger}e_0.
\label{eq:app-Cholesky-basis}
\end{equation}
Since $J_n$ is the leading principal block of $J_{n+1}$, the Cholesky factors are nested principal factors. Consequently, the first $n$ columns produced by this construction do not change when the Krylov space is enlarged. Equation~\eqref{eq:app-Cholesky-basis} therefore constructs precisely the filtration-adapted sequence above. At saturation it reconstructs the full state-metric basis and QFI distribution algebraically from the Hilbert--Schmidt Lanczos data; before saturation it reconstructs the resolved prefix.

One might ask why one cannot simply perform Lanczos directly in the state metric. The operator $\mathcal K_\rho$ is indeed self-adjoint in that metric, so such a construction is mathematically valid. However, its finite-order Galerkin vector is not the projection used in the QFI hierarchy. Let $\widetilde J_{d_0}=[\langle v_i,\mathcal K_\rho v_j\rangle_\rho]_{i,j=0}^{d_0-1}$ be the state-metric Jacobi matrix at saturation. At saturation, the exact coefficients satisfy
\begin{equation}
\widetilde J_{d_0}\boldsymbol\ell=\beta_\rho e_0, \qquad \beta_\rho=\|B\|_\rho.
\label{eq:app-full-state-system}
\end{equation}
At finite order, however, the first $n$ coefficients are not obtained by simply inverting the leading block $\widetilde J_n$. Partitioning
\begin{equation}
\widetilde J_{d_0}
=
\begin{pmatrix}
\widetilde J_n & E_n\\
E_n^\dagger & D_n
\end{pmatrix},
\label{eq:app-state-block}
\end{equation}
eliminating the unresolved components gives
\begin{equation}
\begin{pmatrix}
\ell_0\\
\vdots\\
\ell_{n-1}
\end{pmatrix}
=
\beta_\rho
\left(
\widetilde J_n-E_nD_n^{-1}E_n^\dagger
\right)^{-1}e_0.
\label{eq:app-Schur-coefficients}
\end{equation}
The Schur correction encodes the influence of all deeper Krylov levels. A separately truncated state-metric Galerkin problem would instead give $\widetilde{\boldsymbol\ell}^{(n)}=\beta_\rho\widetilde J_n^{-1}e_0$. In general, the Schur-corrected inverse differs from $\widetilde J_n^{-1}$, because the leading block of an inverse is not the inverse of the leading block. The QFI norm of this separately truncated vector,
\begin{equation}
\|\widetilde L_n\|_\rho^2=\beta_\rho^2e_0^\dagger\widetilde J_n^{-2}e_0,
\label{eq:app-state-squared-inverse}
\end{equation}
is not the cumulative information weight $\sum_{k=0}^{n-1}|\ell_k|^2$. The two coincide only at saturation.

The Hilbert--Schmidt Lanczos matrix $J_n$ and the filtration-adapted state-metric basis therefore play complementary roles: $J_n$ provides the computational QFI hierarchy, while the state-metric basis interprets its completed increments as information weights. A single Hilbert--Schmidt recursion produces the resolved prefix. The complete levelwise decomposition requires saturation; an independently known exact normalization instead supplies a normalized prefix and only the aggregate unresolved remainder.


\section{Spectral Representations and Operator Geometry}
\label{app:spectral-rho}

This appendix collects the spectral and component representations underlying the dual-measure formulation of Sec.~\ref{sec:spectral}. The Hilbert--Schmidt and state-metric measures are supported on the same active spectrum, but they generate two related polynomial systems with different finite-order roles. The component representation of the two metrics and the treatment of rank-deficient states are also given.

Let $\mathcal A=\mathcal K_\rho$, let $u_0=B/\beta_0$, and let $\mathrm d\nu$ be the Hilbert--Schmidt spectral measure defined in Eq.~\eqref{eq:nu-measure}. The spectral theorem gives a unitary map from the Hilbert--Schmidt cyclic space generated by $u_0$ to $L^2(\nu)$ under which
\begin{equation}
 Q(\mathcal A)u_0\longleftrightarrow Q(\lambda), \qquad \mathcal A\longleftrightarrow \lambda.
 \label{eq:app-cyclic-map}
\end{equation}
If $Q_k$ is the polynomial corresponding to the $k$th Hilbert--Schmidt Lanczos vector, then
\begin{equation}
 \int Q_j(\lambda)Q_k(\lambda)\,\mathrm d\nu(\lambda)=\delta_{jk}
\end{equation}
and the three-term recurrence is
\begin{equation}
 \lambda Q_k(\lambda)
 =b_{k+1}Q_{k+1}(\lambda)
 +a_kQ_k(\lambda)+b_kQ_{k-1}(\lambda),
 \label{eq:app-Q-recurrence}
\end{equation}
with $Q_{-1}=0$, $Q_0=1$, and $b_k>0$ before Krylov saturation. The recurrence coefficients form the Jacobi matrix $J_n$ appearing in the Hilbert--Schmidt Lanczos construction.

Let $\zeta_1^{(n)},\ldots,\zeta_n^{(n)}$ be the zeros of $Q_n$. They are real, simple, and lie in the convex hull of $\operatorname{supp}\nu$. They are also the eigenvalues of $J_n$. Zeros of consecutive orthogonal polynomials interlace, so the effective spectral grid is refined as the Krylov dimension grows. If $s_j^{(n)}$ is a normalized eigenvector of $J_n$ with eigenvalue $\zeta_j^{(n)}$, the Gaussian weight is
\begin{equation}
 \omega_j^{(n)}=\left|e_0^\dagger s_j^{(n)}\right|^2.
 \label{eq:app-Gauss-weight-eigenvector}
\end{equation}
Equivalently, it is the Christoffel number
\begin{equation}
 \omega_j^{(n)}
 =\left[
 \sum_{k=0}^{n-1}Q_k\!\left(\zeta_j^{(n)}\right)^2
 \right]^{-1}.
 \label{eq:app-Christoffel-number}
\end{equation}
The $n$-point rule is
\begin{equation}
 \mathcal G_n[f]=\sum_{j=1}^{n}\omega_j^{(n)}f\!\left(\zeta_j^{(n)}\right),
 \label{eq:app-Gauss-rule}
\end{equation}
which is exact for all polynomials of degree at most $2n-1$.

For a sufficiently smooth function on a compact interval containing the support, the Gaussian error can be written in terms of the monic orthogonal polynomial $\pi_n$ as
\begin{equation}
 \int f(\lambda)\,\mathrm d\nu(\lambda)-\mathcal G_n[f]
 =\frac{f^{(2n)}(\xi)}{(2n)!}
 \int\pi_n(\lambda)^2\,\mathrm d\nu(\lambda)
 \label{eq:app-Gauss-error}
\end{equation}
for some $\xi$ in the convex hull of the support. When the support is separated from zero and $f(\lambda)=1/\lambda$, the derivative in Eq.~\eqref{eq:app-Gauss-error} is positive. Hence the Gauss value lies below the exact inverse moment. In the QFI problem,
\begin{equation}
 \mathcal F_n
 =\beta_0^2\mathcal G_n[\lambda^{-1}]
 =\beta_0^2e_0^\dagger J_n^{-1}e_0.
 \label{eq:app-QFI-Gauss}
\end{equation}
The monotone lower-bound property also follows directly from the variational projection and remains valid without invoking the differentiable error formula. At a hard edge, $1/\lambda$ is not smooth on the closed support, and Eq.~\eqref{eq:app-Gauss-error} cannot be used uniformly. The convergence must then be extracted from the endpoint behavior of the measure, as done exactly for the Jacobi class in Appendix~\ref{app:Jacobi-hard-edge}.

The state-metric measure is the Christoffel transform
\begin{equation}
 \mathrm d\mu(\lambda)=\frac{\lambda}{m_1}\,\mathrm d\nu(\lambda).
\end{equation}
Let $P_n$ be orthonormal with respect to $\mu$. When $Q_n(0)\neq0$, the Christoffel formula gives
\begin{equation}
 P_n(\lambda)
 =\mathcal N_n
 \frac{
 Q_{n+1}(\lambda)-\sigma_nQ_n(\lambda)
 }{\lambda},
 \qquad
 \sigma_n=\frac{Q_{n+1}(0)}{Q_n(0)},
 \label{eq:app-Christoffel-polynomials}
\end{equation}
where $\mathcal N_n$ is fixed by normalization in $L^2(\mu)$. The numerator vanishes at the transformation point $\lambda=0$, so the quotient is a polynomial of degree $n$. This formula makes explicit that the two polynomial systems are related rather than independent. Their zeros obey the corresponding Christoffel interlacing relations. At the level of the saturated Jacobi operators, positivity permits a Cholesky factorization
\begin{equation}
 J=R^\dagger R.
\end{equation}
The Jacobi operator associated with the Christoffel-transformed measure is the Darboux partner
\begin{equation}
 \widetilde J=RR^\dagger.
 \label{eq:app-Darboux}
\end{equation}
This is the spectral counterpart of the basis transformation described in Appendix~\ref{app:Krylov-review}. It should not be confused with replacing a full inverse by the inverse of a finite principal block. The state-metric coefficients are
\begin{equation}
 \ell_k=\beta_\rho\int\frac{P_k(\lambda)}{\lambda}
 \mathrm d\mu(\lambda),
\end{equation}
and the dual Krylov theorem gives
\begin{equation}
 \beta_\rho^2\sum_{k=0}^{n-1}
 \left|\int\frac{P_k(\lambda)}{\lambda}
 \mathrm d\mu(\lambda)\right|^2
 =\beta_0^2\mathcal G_n[\lambda^{-1}].
 \label{eq:app-partial-Parseval-Gauss}
\end{equation}
Thus the same finite-order QFI is a partial Parseval sum for the Christoffel-transformed measure and a Gaussian quadrature value for the original source measure.

Turning to the component representation, let $\rho=\sum_a \varrho_a|a\rangle\langle a|$ and define
\begin{equation}
 w_{ab}=\frac{\varrho_a+\varrho_b}{2}.
 \label{eq:app-w-ab}
\end{equation}
In this basis, the state-dependent inner product is
\begin{equation}
 \langle A,B\rangle_\rho
 =\sum_{ab}w_{ab}A_{ab}^*B_{ab}.
 \label{eq:app-state-inner-components}
\end{equation}
The canonical operators $E_{ab}=|a\rangle\langle b|$ satisfy
\begin{equation}
 \mathcal K_\rho(E_{ab})=w_{ab}E_{ab},
 \qquad
 \|E_{ab}\|_\rho^2=w_{ab}.
 \label{eq:app-Eab}
\end{equation}
For unitary encoding,
\begin{equation}
 B_{ab}=i(\varrho_a-\varrho_b)H_{ab},
 \qquad
 L_{ab}=\frac{i(\varrho_a-\varrho_b)}{w_{ab}}H_{ab},
 \label{eq:app-BL-components}
\end{equation}
and the QFI is
\begin{equation}
 \mathcal F
 =\sum_{a,b}\frac{(\varrho_a-\varrho_b)^2}{w_{ab}}|H_{ab}|^2,
 \label{eq:sld-qfi-spectral}
\end{equation}
with terms $\varrho_a+\varrho_b=0$ omitted. The Hilbert--Schmidt and state-metric spectral measures take the explicit forms
\begin{equation}
 d\nu(\lambda)
 =\frac{1}{\beta_0^2}\sum_{ab}|B_{ab}|^2
 \delta(\lambda-w_{ab})d\lambda,
 \label{eq:nu-explicit}
\end{equation}
and
\begin{equation}
 d\mu(\lambda)
 =\frac{1}{\beta_\rho^2}\sum_{ab}w_{ab}|B_{ab}|^2
 \delta(\lambda-w_{ab})d\lambda.
 \label{eq:mu-explicit}
\end{equation}
They satisfy $d\mu=(\lambda/m_1)d\nu$. The state metric contributes exactly one factor of $w_{ab}$; an additional factor would double-count the metric.

For a state-orthonormal basis of the saturated cyclic subspace, the Krylov coefficients are
\begin{equation}
 \ell_k
 =\langle v_k,L\rangle_\rho
 =i\sum_{a,b}(\varrho_a-\varrho_b)v_k^{*ab}H_{ab}.
 \label{eq:app-ell-components}
\end{equation}
The Krylov basis resolves the projector onto the cyclic subspace rather than the identity on the entire operator space. Since $L=\mathcal K_\rho^{-1}B$ belongs to that invariant cyclic subspace, Parseval's identity within the subspace gives
\begin{equation}
 \mathcal F=\|L\|_\rho^2
 =\sum_{k=0}^{d_0-1}|\ell_k|^2.
 \label{eq:app-cyclic-parseval}
\end{equation}
No full-operator-space completeness assumption is required.

Finally, if $\rho$ is not full rank, $\mathcal K_\rho$ is positive semidefinite. Let $\mathcal K_\rho^+$ denote its Moore--Penrose inverse. A finite SLD exists on the active subspace provided
\begin{equation}
 B\in\operatorname{Ran}\mathcal K_\rho.
 \label{eq:app-range}
\end{equation}
Then the minimum-norm SLD is $L=\mathcal K_\rho^+B$, and all formulas hold after restriction to the active cyclic subspace. For unitary encoding, the range condition is automatic because $B_{ab}=0$ whenever $\varrho_a=\varrho_b=0$.

\section{Exact Jacobi Hard-Edge Calculation}
\label{app:Jacobi-hard-edge}

This appendix derives the exact convergence law for the Jacobi hard-edge
family introduced in Sec.~\ref{sec:spectral}. The purpose is not to assume
a universal hard-edge scaling, but to provide an analytically solvable
spectral model in which the relation between the low-energy Liouville
spectrum, the QFI Krylov distribution, and the convergence of the Krylov
hierarchy can be obtained exactly.

The hard-edge regime is characterized by the fact that the support of the
spectral measure reaches the singular point of the QFI resolvent at
$\lambda=0$. In contrast to the gapped case, where the inverse response is
analytic on the spectral support and convergence is exponential, the
presence of low-$\lambda$ Liouville modes produces algebraic convergence in
the Jacobi hard-edge family considered here.

We take the normalized state-metric spectral measure to be
\begin{equation}
\mathrm d\mu_\alpha(\lambda)
=
(\alpha+1)\lambda^\alpha\,\mathrm d\lambda,
\qquad
0\leq\lambda\leq1,
\qquad
\alpha>1.
\label{eq:app-mu-alpha}
\end{equation}
The corresponding QFI is
\begin{equation}
\mathcal F
=
\beta_\rho^2
\int_0^1\frac{\mathrm d\mu_\alpha(\lambda)}{\lambda^2}
=
\beta_\rho^2(\alpha+1)
\int_0^1\lambda^{\alpha-2}\,\mathrm d\lambda
=
\beta_\rho^2\frac{\alpha+1}{\alpha-1}.
\label{eq:app-Falpha}
\end{equation}
For $1<\alpha\leq2$, this finite energy norm does not imply a
Hilbert--Schmidt limiting SLD: using the measure below,
$\|L\|_{\mathrm{HS}}^2$ is proportional to
$\int_0^1\lambda^{-2}\mathrm d\nu_\alpha(\lambda)$ and diverges.  The
continuum expansion is then understood in the state-metric completion of
the cyclic space, while every finite-cutoff formula remains an ordinary
finite-dimensional operator identity.
The associated Hilbert--Schmidt spectral measure is obtained from the
Christoffel relation
\begin{equation}
\mathrm d\mu_\alpha(\lambda)
=
\frac{\lambda}{m_1}\,\mathrm d\nu_\alpha(\lambda),
\qquad
m_1=\int_0^1\lambda\,\mathrm d\nu_\alpha(\lambda)
=\frac{\alpha}{\alpha+1},
\end{equation}
and is therefore
\begin{equation}
\mathrm d\nu_\alpha(\lambda)
=
\alpha\lambda^{\alpha-1}\,\mathrm d\lambda .
\label{eq:app-nu-alpha}
\end{equation}
Thus the two measures describe the same Liouville spectrum with different
weightings: the state-metric measure describes the SLD geometry, whereas
the Hilbert--Schmidt measure describes the computational Lanczos
representation.

The normalized spectral distribution of the QFI is
\begin{equation}
\mathrm d\pi_{\mathcal F}(\lambda)
=
\frac{\beta_\rho^2}{\mathcal F}
\frac{\mathrm d\mu_\alpha(\lambda)}{\lambda^2}
=
(\alpha-1)\lambda^{\alpha-2}\,\mathrm d\lambda .
\label{eq:app-qfi-measure}
\end{equation}
Hence $\alpha$ controls not only the convergence rate but also the
distribution of metrological sensitivity. For $1<\alpha<2$, the QFI
density diverges toward the edge, although the mass of a shrinking interval
$[0,\delta]$ is $\delta^{\alpha-1}$ and tends to zero. For $\alpha=2$,
the QFI contribution is uniform, whereas for $\alpha>2$ its edge density
is suppressed.

The orthonormal polynomials with respect to
$\mathrm d\mu_\alpha$ are~\cite{Szego1975,DLMF18}
\begin{equation}
P_k(\lambda)
=
\sqrt{\frac{2k+\alpha+1}{\alpha+1}}\,
P_k^{(0,\alpha)}(2\lambda-1),
\label{eq:app-Jacobi-polys}
\end{equation}
where $P_k^{(0,\alpha)}$ denotes the Jacobi polynomial.

Using the Rodrigues representation,
\begin{equation}
P_k^{(0,\alpha)}(2\lambda-1)
=
\frac{(-1)^k}{k!}\lambda^{-\alpha}
\frac{\mathrm d^k}{\mathrm d\lambda^k}
\left[
\lambda^{k+\alpha}(1-\lambda)^k
\right],
\end{equation}
one finds
\begin{align}
&\int_0^1
\lambda^{\alpha-1}
P_k^{(0,\alpha)}(2\lambda-1)\,\mathrm d\lambda
\cr
&=
\frac{(-1)^k}{k!}
\int_0^1
\lambda^{-1}
\frac{\mathrm d^k}{\mathrm d\lambda^k}
\left[
\lambda^{k+\alpha}(1-\lambda)^k
\right]\mathrm d\lambda
\nonumber\\
&=
(-1)^k \mathrm B(\alpha,k+1).
\label{eq:app-Jacobi-integral}
\end{align}
The boundary terms vanish because the differentiated factor contains
$\lambda^{k+\alpha}(1-\lambda)^k$, whose derivatives of order less than
$k$ vanish at both endpoints for $\alpha>1$.

The SLD coefficients in the filtration-adapted state-metric orthonormal basis are therefore
\begin{equation}
\frac{\ell_k}{\beta_\rho}
=
(-1)^k
\sqrt{(\alpha+1)(2k+\alpha+1)}
\,\mathrm B(\alpha,k+1).
\label{eq:app-ell-Jacobi}
\end{equation}
Using
\begin{equation}
\mathrm B(\alpha,k+1)
=
\frac{\Gamma(\alpha)\Gamma(k+1)}
{\Gamma(k+\alpha+1)},
\end{equation}
the corresponding normalized Krylov weights
\begin{equation}
p_k=\frac{|\ell_k|^2}{\mathcal F}
\end{equation}
are
\begin{equation}
p_k
=
(\alpha-1)(2k+\alpha+1)
\left[
\frac{\Gamma(\alpha)\Gamma(k+1)}
{\Gamma(k+\alpha+1)}
\right]^2.
\label{eq:app-pk-Jacobi}
\end{equation}

Define the tail function
\begin{equation}
S_n
=
\left[
\frac{\Gamma(\alpha)\Gamma(n+1)}
{\Gamma(n+\alpha)}
\right]^2.
\label{eq:app-Sn}
\end{equation}
A direct use of $\Gamma(z+1)=z\Gamma(z)$ gives
\begin{equation}
S_k-S_{k+1}=p_k.
\label{eq:app-telescoping}
\end{equation}
Since $S_n\to0$ for $\alpha>1$,
\begin{equation}
\sum_{k=n}^{\infty}p_k=S_n.
\label{eq:app-tail}
\end{equation}
Because
\begin{equation}
1-\frac{\mathcal F_n}{\mathcal F}
=
\sum_{k=n}^{\infty}p_k,
\end{equation}
the finite-order QFI error is exactly
\begin{equation}
1-\frac{\mathcal F_n}{\mathcal F}
=
\left[
\frac{\Gamma(\alpha)\Gamma(n+1)}
{\Gamma(n+\alpha)}
\right]^2.
\label{eq:app-exact-Jacobi-error}
\end{equation}

The gamma-ratio asymptotic
\begin{equation}
\frac{\Gamma(n+1)}{\Gamma(n+\alpha)}
\sim n^{1-\alpha}
\end{equation}
therefore gives
\begin{equation}
1-\frac{\mathcal F_n}{\mathcal F}
\sim
\Gamma(\alpha)^2 n^{-2(\alpha-1)}.
\label{eq:app-Jacobi-asymptotic}
\end{equation}
Thus the Jacobi hard edge provides an exact example of algebraic Krylov
convergence, with an exponent determined by the edge parameter $\alpha$.

Similarly, Eq.~\eqref{eq:app-ell-Jacobi} gives
\begin{equation}
|\ell_k|
\sim
\beta_\rho
\sqrt{2(\alpha+1)}\,\Gamma(\alpha)\,
k^{1/2-\alpha}.
\label{eq:app-ell-asymptotic}
\end{equation}
The coefficients therefore decay algebraically even though the total QFI
remains finite for every $\alpha>1$.

The tail function $S_n$ also determines the information depth. With
\begin{equation}
{\cal D}=\sum_{k=0}^{\infty} k p_k,
\end{equation}
the standard tail-sum identity gives
\begin{equation}
{\cal D}_\alpha
=
\sum_{n=1}^{\infty}S_n.
\label{eq:app-Dalpha-tail}
\end{equation}
Using
\begin{equation}
S_n
=
\left[
\frac{(1)_n}{(\alpha)_n}
\right]^2,
\end{equation}
this can be written as
\begin{equation}
{\cal D}_\alpha
=
{}_3F_2(1,1,1;\alpha,\alpha;1)-1,
\label{eq:app-Dalpha-hypergeometric}
\end{equation}
whenever the series converges.

Since
\begin{equation}
S_n
\sim
\Gamma(\alpha)^2 n^{-2(\alpha-1)},
\end{equation}
the information depth is finite if and only if
\begin{equation}
\alpha>\frac{3}{2}.
\label{eq:app-D-threshold}
\end{equation}
For
\begin{equation}
1<\alpha\leq\frac32,
\end{equation}
the total QFI remains finite, but the mean Krylov depth diverges: at
$\alpha=3/2$ the divergence is logarithmic, whereas for
$1<\alpha<3/2$ it is algebraic. Thus finite QFI, which requires only
$\alpha>1$, does not imply finite information depth.

As a useful check, for $\alpha=2$ the exact tail is
\begin{equation}
S_n=\frac{1}{(n+1)^2},
\qquad
D_2
=
\sum_{n=1}^{\infty}\frac{1}{(n+1)^2}
=
\frac{\pi^2}{6}-1.
\label{eq:app-Dalpha-two}
\end{equation}
This exact algebraic law illustrates how low-$\lambda$ Liouville modes
slow the resolution of the QFI despite the finiteness of the total
susceptibility.

For completeness, the boundary case $\alpha=1$ has
\begin{equation}
\mathrm d\mu_1(\lambda)
=
2\lambda\,\mathrm d\lambda,
\qquad
0\leq\lambda\leq1,
\label{eq:app-mu-marginal}
\end{equation}
with associated Hilbert--Schmidt measure
\begin{equation}
\mathrm d\nu_1(\lambda)=\mathrm d\lambda.
\label{eq:app-nu-marginal}
\end{equation}
The continuum QFI is then
\begin{equation}
\mathcal F
=
2\beta_\rho^2
\int_0^1\frac{\mathrm d\lambda}{\lambda},
\end{equation}
and therefore diverges logarithmically at the hard edge.

A finite system necessarily introduces an infrared regularization. For
example, with $\varepsilon=\lambda_{\min}>0$, define the normalized
regulated measure
\begin{equation}
 \mathrm d\mu_{1,\varepsilon}(\lambda)
 =\frac{2\lambda}{1-\varepsilon^2}
 \mathbf 1_{[\varepsilon,1]}(\lambda)\,\mathrm d\lambda.
 \label{eq:app-mu-marginal-regulated}
\end{equation}
Its regulated susceptibility is
\begin{equation}
\mathcal F(\varepsilon)
=\frac{2\beta_{\rho,\varepsilon}^2}{1-\varepsilon^2}
\ln\frac{1}{\varepsilon}
\sim 2\beta_{\rho,\varepsilon}^2\ln\frac{1}{\varepsilon}.
\label{eq:app-F-marginal-cutoff}
\end{equation}
The leading logarithmic divergence is therefore robust, although the
finite-size regularization and the associated finite-order Krylov
convergence depend on how the infrared cutoff is implemented.

In particular, once the spectral support is restricted to
$[\varepsilon,1]$, the resulting measure is no longer the Jacobi
measure on the full interval $[0,1]$. Consequently, the orthogonal
polynomials and finite-order Krylov coefficients of the regulated problem
are not, in general, the uncut $\alpha=1$ Jacobi polynomials used above.
We therefore do not infer a universal finite-$n$ convergence law for the
marginal case from the uncut Jacobi basis. The $\alpha=1$ boundary is
instead excluded from the cutoff-independent Jacobi convergence theorem
derived above.

It is worth noting that the exact result above applies to the Jacobi family
\begin{equation}
\mathrm d\mu_\alpha(\lambda)
=
(\alpha+1)\lambda^\alpha\,\mathrm d\lambda
\end{equation}
on $[0,1]$. It should not be interpreted as a universal convergence law
for every spectral measure having the same local edge exponent. More
generally, measures of the form
\begin{equation}
\mathrm d\mu(\lambda)
=
\lambda^\alpha h(\lambda)\,\mathrm d\lambda
\end{equation}
require further assumptions on the regularity and global properties of
$h(\lambda)$ before analogous convergence laws can be established.
Accordingly, the Jacobi calculation should be viewed as an exactly
solvable hard-edge benchmark that demonstrates how a low-energy spectral
edge can produce algebraic Krylov convergence, rather than as a universal
hard-edge theorem.


\bibliography{literature}

\end{document}